\documentclass[aps,showpacs,twocolumn,preprintnumbers,amsmath,amssymb]{revtex4}
\usepackage{graphicx}% Include figure files
\usepackage{bm}% bold math

\def\bk{ \bm{k} }
\def\bK{ \bm{K} }

\def\bA{ \bm{A} }
\def\bB{ \bm{B} }

\def\re{ \,\mathrm{Re}\, }
\def\blam{ \bm{\lambda} }

\begin{document}

\title{Paramagnetic properties of non-centrosymmetric superconductors:\\
Application to CePt$_3$Si}

\author{K.~V.~Samokhin}

\affiliation{Department of Physics, Brock University,
St.Catharines, Ontario, Canada L2S 3A1}
\date{\today}

\begin{abstract}
In a non-centrosymmetric crystal, the Zeeman interaction of the
band electrons with an external magnetic field is highly
anisotropic in the momentum space, vanishing along some
high-symmetry planes. One of the consequences is that the
paramagnetic susceptibility in superconductors without inversion
symmetry, such as CePt$_3$Si, shows an unusual temperature
dependence.
\end{abstract}

\pacs{74.20.Rp, 74.25.Ha, 74.70.Tx}

\maketitle

Electronic spin susceptibility measurements in the superconducting
state provide one of the most useful tools for the identification
of the pairing symmetry. Experimentally, the susceptibility
$\chi(T)$ is measured by the Knight shift in the nuclear magnetic
resonance frequency due to the hyperfine interaction of the
conduction electrons with the nuclear magnetic moments. If the
superconducting pairing occurs in the spin-singlet channel, then
the Cooper pairs have spin $S=0$ and do not contribute to the
magnetization $\bm{M}$ of the system, which is therefore entirely
determined by the thermally excited quasiparticles. For a fully
gapped order parameter one would see an exponentially decreasing
$\chi(T)$ at low temperatures. In contrast, if the gap has zeros
at the Fermi surface, then $\chi(T)\propto T^2$ for isolated point
nodes, or $\chi(T)\propto T$ for line nodes. In the triplet case,
when the Cooper pairs have spin $S=1$ and the order parameter is a
vector $\bm{d}(\bk)$ in the spin space (see, e.g., Ref.
\cite{Book}), the susceptibility depends on the mutual orientation
of $\bm{d}$ and the external magnetic field $\bB$. If
$\bB\parallel\bm{d}$, then the Cooper pairs do not contribute to
$\bm{M}$, and $\chi(T)$ has the same temperature dependence as in
the singlet case. If $\bB\perp\bm{d}$, then both the pairs and the
excitations contribute to $\bm{M}$, so that $\chi(T)=\chi_n$ --
the normal-state susceptibility. The observation of a flat Knight
shift in Sr$_2$RuO$_4$ for the field in the basal $xy$ plane
\cite{SRO-NMR}, has been used as a proof of a spin-triplet pairing
with $\bm{d}\parallel\hat z$.

The theoretical picture described above is valid only if the
superconducting crystal has an inversion center. Although this is
the case in the majority of superconductors, some exceptions have
been known since 1960's \cite{early examples}. More recently, it
was pointed out in Ref. \cite{GR01} that the surface
superconductors, e.g. Na-doped WO$_3$ \cite{WO3}, are
intrinsically non-centrosymmetric simply because the two sides of
the surface layer are manifestly non-equivalent. As for the bulk
materials, the latest examples are CePt$_3$Si \cite{exp CePtSi}
and UIr \cite{exp UIr}. Different models of superconductivity in
CePt$_3$Si have been proposed in Refs. \cite{SZB04,FAKS04}. In
this Letter, we calculate the suppression of the critical
temperature and the paramagnetic susceptibility in superconductors
without an inversion center. We focus on the tetragonal symmetry
relevant for CePt$_3$Si, and show in particular that the
anisotropy of the temperature dependence of the susceptibility
tensor $\chi_{ij}(T)$ is strikingly different from the
centrosymmetric case.

The spin susceptibility in two-dimensional non-centrosymmetric
superconductors has been previously studied in Refs.
\cite{BGR76,Edelstein,GR01,Yip02}, where the inversion symmetry
breaking in the presence of a non-zero spin-orbit (SO) coupling
was introduced using the Rashba model \cite{Rashba}. Very
recently, a three-dimensional generalization of the Rashba model
was applied to CePt$_3$Si in Ref. \cite{FAS04}. Treating the SO
band splitting as a perturbation, it was found that the order
parameter becomes a mixture of a spin-singlet (even in $\bk$) and
a spin-triplet (odd in $\bk$) components, which gives rise to a
non-zero residual susceptibility at $T=0$. In this Letter, we use
a different approach based on the effective single-band
Hamiltonian, which works for any crystal symmetry and arbitrary
strength of the SO coupling. Guided by the fact that the SO band
splitting is usually large compared to the superconducting energy
scales, we consider both the Cooper pairing and the magnetic
response independently in different bands. In contrast to the
previous works, we construct the pairing interaction using the
exact band states and explicitly take into account that all the
pairing channels but one are suppressed \cite{And84}.

The starting point of our analysis is the observation that in a
crystal lacking an inversion center the electron bands are
non-degenerate almost everywhere, except along some high-symmetry
lines in the Brillouin zone. Indeed, without the inversion
operation $I$, one cannot construct two orthogonal degenerate
Bloch states at the same $\bk$. At zero SO coupling there is an
additional symmetry in the system -- the invariance with respect
to arbitrary spin rotations, which restores the two-fold spin
degeneracy of the bands. Here we assume that the SO coupling is
sufficiently strong, so that the bands are well split. The results
of Ref. \cite{SZB04} show that this is indeed the case in
CePt$_3$Si, where the SO band splitting can be as large as 50-200
meV depending on the band. Assuming that there is no disorder in
the crystal, the Bloch wave vector $\bk$ is a good quantum number
in zero field. The free electron Hamiltonian for a non-degenerate
band can be written as
$H_0=\sum_{\bk}\epsilon(\bk)c^\dagger_{\bk}c_{\bk}$, where
$\sum_{\bk}$ stands for the integration over the first Brillouin
zone, and $\epsilon(\bk)$ is the quasiparticle dispersion, which
takes into account the periodic lattice potential and the SO
interaction.

If the time-reversal symmetry is not broken in the normal phase,
then the states $|\bk\rangle$ and $K|\bk\rangle\sim|-\bk\rangle$
are degenerate because of the Kramers theorem \cite{AFM comment}.
It is the coupling between those states that leads to the
formation of the Cooper pairs and the superconductivity in the
system. The large band splitting strongly suppresses the pairing
of electrons from different bands. Then, considering just one band
(i.e. neglecting the inter-band pair scattering) and assuming that
the interaction has a generalized Bardeen-Cooper-Schrieffer (BCS)
form, we have
\begin{equation}
\label{H BCS}
    H_{int}=\frac{1}{2}\sum\limits_{\bk,\bk'}
    V(\bk,\bk')c^\dagger_{\bk}c^\dagger_{-\bk}c_{-\bk'}c_{\bk'}.
\end{equation}
The pairing potential can be written as $V(\bk,\bk')=\tilde
V(\bk,\bk')t(\bk)t^*(\bk')$, where $\tilde
V(\bk,\bk')=-V_\Gamma\sum_a\phi_a(\bk)\phi_a^*(\bk')$ is the part
that transforms according to an irreducible representation
$\Gamma$ of the normal-state point group $G$, $\phi_a(\bk)$ are
the scalar basis functions of $\Gamma$, which are are nonzero only
inside the energy shell of width $\epsilon_c$ near the Fermi
surface, $V_\Gamma>0$ is the coupling constant, and
$t(\bk)=-t(-\bk)$ are non-trivial phase factors in
$K|\bk\rangle=t(\bk)|-\bk\rangle$ \cite{SC04}. Although
anti-commutation of fermionic operators dictates that the
mean-field order parameter
$\Delta(\bk)=t(\bk)\sum_a\eta_a\phi_a(\bk)$ is odd in $\bk$
\cite{SZB04}, its nodal structure is determined by the basis
functions $\phi_a(\bk)$, which should be even because of the
presence of $t(\bk)$ \cite{correct}. The focus of this article is
on CePt$_3$Si, which has a non-centrosymmetric tetragonal crystal
lattice described by the point group $G=\mathbf{C}_{4v}$. This
group is generated by the rotations $C_{4z}$ about the $z$ axis by
an angle $\pi/2$ and the reflections $\sigma_x$ in the vertical
plane $(100)$, and has five irreducible representations: four
one-dimensional (1D): $A_1$, $A_2$, $B_1$, $B_2$, and one
two-dimensional $E$ \cite{LL3}. Here are the examples of the even
basis functions: $\phi_{A_1}\propto k_x^2+k_y^2+ck_z^2$,
$\phi_{A_2}\propto k_xk_y(k_x^2-k_y^2)$, $\phi_{B_1}\propto
k_x^2-k_y^2$, $\phi_{B_2}\propto k_xk_y$, and
$(\phi_{E,1},\phi_{E,2})\propto (k_xk_z,k_yk_z)$. We consider a
small sample, of a dimension $d\leq\xi<\delta$, where $\xi$ is the
superconducting correlation length and $\delta$ is the London
penetration depth, which allows us to neglect the spatial
variations of both the order parameter components $\eta_a$ and the
magnetic field.

Let us now turn on a uniform stationary magnetic field
$\bB=\mathrm{curl}\,\bA$. Assuming that the pairing interaction is
field-independent, $\bB$ can only affect the system through its
coupling to the band states. At $\bB\neq 0$, the band dispersion
function $\epsilon(\bk)$ is replaced by an effective band
Hamiltonian in the momentum space, which can be represented as a
power series in $\bB$: $\epsilon(\bk)\to{\cal
E}(\bk,\bB)=\epsilon(\bK)+B_i\epsilon_{1,i}(\bK)+...$, where
$\bK=\bk+(e/\hbar c)\bA(i\bm{\nabla}_{\bk})$ because of the
requirements of gauge invariance \cite{H-eff}. The expansion
coefficients must satisfy certain symmetry-imposed conditions, in
particular the zero-field band dispersion $\epsilon(\bk)$ must be
invariant under all operations from $G$. In addition, at $\bB\neq
0$ the Hamiltonian is invariant with respect to time reversal $K$
only if the sign of $\bB$ (and of $\bA$) is also changed, which
imposes the following constraint on the function ${\cal E}$:
$K^\dagger{\cal E}(-\bB)K={\cal E}(\bB)$.

In the analysis of the ``paramagnetic'' properties of
superconductors, the orbital effect of the field is neglected,
which is achieved by putting $\bA=0$ in the effective band
Hamiltonian. Then, for a two-fold degenerate band in the presence
of inversion symmetry, ${\cal E}$ is a $2\times 2$ matrix, and the
coupling to the magnetic field is described by a familiar Zeeman
term: ${\cal
E}_{\alpha\beta}(\bk,\bB)=\epsilon(\bk)\delta_{\alpha\beta}-B_i\mu_{ij}(\bk)
\sigma_{j,\alpha\beta}$, with $\mu_{ij}(\bk)=\mu_{ij}(-\bk)$ being
the tensor generalization of the Bohr magneton $\mu_B$ for the
case of band electrons. The indices $\alpha\beta$ here are
pseudospin indices \cite{UR85}. The Zeeman interaction splits the
energies of the electrons forming the Cooper pairs and gives rise
to the paramagnetic suppression of superconductivity \cite{CC62}.

If the inversion symmetry is absent and the bands are
non-degenerate, then the Zeeman term should be modified. The
effective single-band Hamiltonian in the external field can be
written as
\begin{equation}
\label{H 0}
    H_0=\sum\limits_{\bk}\left[\epsilon(\bk)-\bB\blam(\bk)\right]
    c^\dagger_{\bk} c_{\bk},
\end{equation}
which is markedly different from the centrosymmetric case. Here
$\blam(\bk)$ is a real pseudovector, which satisfies the
conditions $(g\blam)(g^{-1}\bk)=\blam(\bk)$, where $g$ is any
operation from the point group $G$. Because of the time-reversal
symmetry, we also have $\epsilon(-\bk)=\epsilon(\bk)$ and
$\blam(-\bk)=-\blam(\bk)$, but ${\cal E}(-\bk,\bB)\neq{\cal
E}(\bk,\bB)$.

Explicit expressions for $\blam(\bk)$ can only be obtained in some
simple models. For example, in an isotropic two-dimensional
electron gas in the $xy$ plane with $G=\mathbf{C}_{\infty v}$, the
combined effect of the SO coupling and the lack of inversion
symmetry is described by an additional (Rashba) term in the
single-particle Hamiltonian:
$H_{SO}=\gamma\sum_{\bk}\bm{n}\cdot(\bm{\sigma}_{\sigma\sigma'}
\times\bk)\,a^\dagger_{\bk\sigma}a_{\bk\sigma'}$, where $\bm{n}$
is the normal vector to the plane \cite{Rashba}. Diagonalization
of the Hamiltonian in zero field gives two non-degenerate Rashba
bands: $\epsilon_\pm(\bk)=\epsilon_0(\bk)\pm\gamma|\bk|$. At
finite field, adding a usual Zeeman term
$-\mu_B(\bB\cdot\bm{\sigma})$, and expanding the eigenlavues of
the Hamiltonian in powers of $\bB$, we obtain ${\cal
E}_\pm(\bk,\bB)=\epsilon_\pm(\bk)-\blam_\pm(\bk)\bB+O(B^2)$, where
$\blam_\pm(\bk)=\pm\mu_B(\bk\times\bm{n})/|\bk|$. Thus the
coupling of the Rashba bands with the field is highly anisotropic,
in particular it vanishes for $\bB\parallel\bm{n}$.

While a microscopic derivation of the effective single-band
Hamiltonian (\ref{H 0}) in more realistic systems can be done, at
least in principle, using the procedures described in Refs.
\cite{H-eff}, it suffices for our purposes to work with a
phenomenological expression for $\blam(\bk)$, which is compatible
with all the symmetry constraints. We need an expression for
$\blam(\bk)$, which satisfies (i) $\blam(-\bk)=-\blam(\bk)$, (ii)
$(C_{4z}\blam)(C_{4z}^{-1}\bk)=\blam(\bk)$, and (iii)
$(\sigma_x\blam)(\sigma_x^{-1}\bk)=\blam(\bk)$ (since $\blam$ is a
pseudovector, we have $\sigma_x\blam\equiv
IC_{2x}\blam=C_{2x}\blam$, where $C_{2x}$ is a rotation by an
angle $\pi$ about the $x$ axis). It is straightforward to check
that the general expression for $\blam(\bk)$ is given by
\begin{equation}
\label{lambda CPS}
    \blam(\bk)=\tilde\phi_{E,2}(\bk)\hat x-
    \tilde\phi_{E,1}(\bk)\hat y+\tilde\phi_{A_2}(\bk)\hat z,
\end{equation}
where $\tilde\phi_{E,1(2)}(\bk)$ and $\tilde\phi_{A_2}(\bk)$ are
real odd functions, which transform according to the
representations $E$ and $A_2$ respectively, e.g.
$\tilde\phi_{A_2}\propto k_xk_yk_z(k_x^2-k_y^2)$ and
$(\tilde\phi_{E,1},\tilde\phi_{E,2})\propto (k_x,k_y)$.  We see
that $\lambda_z=0$ along the five nodal planes of the $A_2$
representation, while $\lambda_x=\lambda_y=0$ along the $z$ axis.

Now, we calculate the free energy ${\cal F}$ for the Hamiltonian
$H=H_0+H_{int}$, defined by Eqs. (\ref{H 0}) and (\ref{H BCS}). We
use the effective field theory in terms of the bosonic Matsubara
fields $\eta_a(\tau)$, which can be introduced in a standard
fashion by decoupling the pairing interaction (\ref{H BCS}). The
effective action for a uniform stationary order parameter in the
mean-field approximation reads
\begin{equation}
\label{S eff}
    S_{eff}=\frac{\beta{\cal V}}{2V_\Gamma}\sum\limits_a|\eta_a|^2
    -\frac{1}{2}\mathrm{Tr}\ln G^{-1},
\end{equation}
where
\begin{eqnarray}
\label{G inverse}
    &&G^{-1}(\bk,\omega_n)\nonumber\\
    &&=\left(%
    \begin{array}{cc}
    i\omega_n-\epsilon(\bk)+\bB\blam(\bk) & -\Delta(\bk) \\
    -\Delta^*(\bk) & i\omega_n+\epsilon(\bk)+\bB\blam(\bk) \\
    \end{array}%
    \right)\qquad
\end{eqnarray}
is the $2\times 2$ inverse Gor'kov Green's function,
$\omega_n=(2n+1)\pi T$ is the fermionic Matsubara frequency, and
${\cal V}$ is the system volume. The mean-field free energy is
related to the saddle-point action (\ref{S eff}): ${\cal
F}=(1/\beta)S_{eff}$, and the magnetization density is
$\bm{M}=-{\cal V}^{-1}(\partial{\cal F}/\partial\bB)$. The saddle
point condition yields the self-consistency equation
\begin{equation}
\label{gap eq}
    \frac{1}{V_\Gamma}\eta_a+T\sum\limits_n\sum\limits_{\bk}t^*(\bk)\phi_a^*(\bk)G_{12}(\bk,\omega_n)=0,
\end{equation}
which determines the temperature and field dependence of the order
parameter components. Substituting here the Green's function
(\ref{G inverse}), one can see that the phase factors $t(\bk)$
drop out of the gap equation.

Let us first find how the critical temperature is suppressed by
the field. The equation for $T_c(\bB)$ is obtained by linearizing
Eq. (\ref{gap eq}) and can be written in the form
$\det||K_{ab}||=0$, where
\begin{eqnarray}
\label{Kab}
    &&K_{ab}=\left[\ln\frac{T_{c0}}{T_c}+\psi\left(\frac{1}{2}\right)\right]\delta_{ab}\\
    &&\qquad-\left\langle\phi_a^*(\bk)\phi_b(\bk)\re\psi\left(\frac{1}{2}-
    i\frac{\bB\blam(\bk)}{2\pi
    T_c}\right)\right\rangle_{FS},\nonumber
\end{eqnarray}
where $\psi(x)$ is the digamma function, $T_{c0}\simeq
1.13\epsilon_c\exp(1/N_FV_\Gamma)$ is the critical temperature in
zero field ($N_F$ is the density of states at the Fermi level),
and the angular brackets denote the average over the Fermi
surface. Next, differentiating the effective action (\ref{S eff})
with respect to $\bB$, we find the magnetization density
\begin{equation}
\label{M}
    \bm{M}=\frac{1}{2}T\sum\limits_n\sum\limits_{\bk}\blam(\bk)\left[
    G_{11}(\bk,\omega_n)+G_{22}(\bk,\omega_n)\right].
\end{equation}
The uniform susceptibility tensor is defined in the usual manner
as $\chi_{ij}=\partial M_i/\partial B_j|_{\bB=0}$. One can easily
check using Eq. (\ref{gap eq}) that the corrections to the order
parameter components $\eta_a$ in a weak magnetic field are
quadratic in $\bB$, which means that in the calculation of
$\chi_{ij}$ one can neglect the field dependence of $\Delta(\bk)$,
to obtain
\begin{equation}
\label{chi ij}
    \chi_{ij}(T)=\frac{1}{4T}\sum\limits_{\bk}
    \frac{\lambda_i(\bk)\lambda_j(\bk)}{\cosh^2[E(\bk)/2T]},
\end{equation}
where $E(\bk)=\sqrt{\epsilon^2(\bk)+|\Delta(\bk)|^2}$.

We now apply the general theory to CePt$_3$Si. The pseudovector
$\blam(\bk)$ in this case is given by Eq. (\ref{lambda CPS}).
Since all three components of $\blam$ are in general non-zero, we
expect the superconducting critical temperature to be suppressed
for all orientations of the magnetic field. According to Eq.
(\ref{Kab}), the magnitude of the suppression depends on many
factors: the shape of the Fermi surface, the symmetry of the order
parameter, and also the explicit form of the functions
$\tilde\phi$ in Eq. (\ref{lambda CPS}). The Fermi surface of
CePt$_3$Si is quite complicated and consists of six sheets
\cite{SZB04}. It is not known which one (or ones) of them are
superconducting. The order parameter symmetry is not known either,
although the observation of a linear $T$-dependence of the
specific heat \cite{exp CePtSi} probably indicates that the order
parameter has lines of nodes. In view of all this uncertainty, it
seems to be premature to discuss the paramagnetic suppression in
CePt$_3$Si quantitatively.

More interesting qualitative conclusions can be drawn from the
analysis of the susceptibility $\chi_{ij}(T)$. It follows from Eq.
(\ref{chi ij}) that the main contribution to the susceptibility
tensor at low temperatures comes from the thermally excited nodal
quasiparticles. If the order parameter $\Delta(\bk)$ has no zeros
at the Fermi surface (e.g. for the $A_1$ representation), the
susceptibilities are exponentially small in all directions. Let us
consider the 1D order parameters corresponding to the
representations $A_2$, $B_1$, or $B_2$, for which the lines of
nodes are symmetry-imposed. In this case, the components of the
susceptibility tensor are given by
$\chi_{xx}=\chi_{yy}=\chi_\parallel$ and $\chi_{zz}=\chi_\perp$,
where
\begin{eqnarray}
\label{chi parall}
    &&\chi_\parallel(T)=\frac{1}{8T}\sum\limits_{\bk}
    \frac{\tilde\phi_{E,1}^2(\bk)+\tilde\phi_{E,2}^2(\bk)}{\cosh^2[E(\bk)/2T]},\\
\label{chi perp}
    &&\chi_\perp(T)=\frac{1}{4T}\sum\limits_{\bk}
    \frac{\tilde\phi_{A_2}^2(\bk)}{\cosh^2[E(\bk)/2T]}.
\end{eqnarray}
It is straightforward to show that for the field in the basal
plane, $\chi_\parallel(T)\propto T$, since
$\tilde\phi_{E,1}^2(\bk)+\tilde\phi_{E,2}^2(\bk)$ is in general
non-zero everywhere, except the poles of the Fermi surface. This
behavior is characteristic of the systems with lines of nodes. In
contrast, for the field orientation along the $z$ axis, we have
from Eq. (\ref{chi perp}) $\chi_\perp(T)\propto T^3$, since
$\tilde\phi_{A_2}(\bk)$ vanishes at the nodal lines for all the 1D
order parameters. Such temperature dependence is never seen in the
centrosymmetric case. For a two-component order parameter
$\Delta(\bk)\sim\eta_1k_xk_z+\eta_2k_yk_z$ with a horizontal line
of zeros at $k_z=0$, one also has a $T^3$-behaviour of
$\chi_\perp(T)$. On the other hand, the anisotropy and the
temperature dependence of $\chi_\parallel(T)$ is determined by
$(\eta_1,\eta_2)$.

We would like to note that, in contrast to Refs.
\cite{GR01,FAS04}, our approach does not yield a finite value of
the susceptibility at $T=0$. The explanation is that the residual
susceptibility in the systems described by the Rashba model comes
from the interband transitions between the Rashba bands (Van Vleck
susceptibility), which are not affected by the transition into the
superconducting state \cite{Yip02}. In our model, the
quasiparticles in each band respond to the external field
independently of the other bands. If to take the interband
transitions into account, then the observed susceptibility will be
$\chi_{tot}(T)=\chi_{0}+\chi(T)$, where the first term is the
temperature-independent background that comes, e.g., from the Van
Vleck processes, and the second term is the single-band
contribution (\ref{chi ij}). Other mechanisms that might affect
the residual susceptibility include the contributions from the
unpaired sheets of the Fermi surface, or spin-reversing scattering
at impurities or surface imperfections \cite{residual chi}.

In conclusion, we have shown that the paramagnetic responses of
superconductors with and without inversion center are
qualitatively different. The most important feature is that, in
the latter case, the coupling of the non-degenerate band electrons
with the external field is strongly momentum-dependent and
vanishes, for symmetry reasons, along some high-symmetry planes in
the Brillouin zone. This results in a high anisotropy of the
susceptibility in the superconducting state with lines of nodes,
from $\chi_\parallel(T)\propto T$ to $\chi_\perp(T)\propto T^3$,
which can be used as a clear-cut experimental test of our theory.

This work was supported by the Natural Sciences and Engineering
Research Council of Canada.


\begin{thebibliography}{99}
\bibitem{correct}
The gap symmetry was analyzed in Ref. \cite{SZB04} using the odd
basis functions $\phi_a(\bk)$. This has been corrected in Ref.
\cite{SC04}.
\end{thebibliography}
\end{document}